# THE TWO POMERONS [*]

P V Landshoff

DAMTP, University of Cambridge

## ABSTRACT

A wide variety of experimental data in hadronic reactions are described in terms of the exchange of an object called the soft pomeron. It is nonperturbative in origin and is perhaps associated with glueballs. The HERA measurements of $\nu W_2$ at small $x$ suggest the possibility that there is also a hard pomeron, which is perturbative. I review the experimental and theoretical methods that are used to explore the two pomerons and emphasise the problems that remain to be solved.

## 1. INTRODUCTION

Throughout the 1960's, most high energy physicists worked on Regge theory, which successfully described a large quantity of hadronic scattering data. It remains one of the great truths of our subject. There was rather little work on Regge theory in the 1970's and 1980's, but interest has now been strongly revived by the HERA measurements of $\nu W_2$ at small $x$.

Regge theory describes forces in terms of exchanges of known particles, and also of new objects called pomerons. There seem to be two pomerons:

> **The soft pomeron**: this is nonperturbative, and so it is very difficult to derive from QCD, but its properties are well known from a huge amount of data that have accumulated over 35 years.

> **The hard pomeron**: this is described by the perturbative BFKL equation of QCD[1] but as yet is supported by little or no experimental data.

In this lecture, I shall outline both the data and the theory. I shall emphasise the problems that remain to be solved. In particular I shall ask, if there are two pomerons, how do they relate to each other?

Regge theory has a sound mathematical basis, namely the theory of complex angular momentum, and a proper account would begin with this. It is well described in the book by Collins[2], which also describes the large amount of early phenomenology. Fortunately, it is not necessary to understand complex angular momentum to use Regge theory. I shall begin my account with total cross section data.

## 2. TOTAL CROSS SECTIONS

Figure 1a shows data for the $pp$ and $\bar p p$ total cross sections, over a very wide range of energies. As indicated in the figure, the curves are each a fit[3] in terms of just two powers. I shall explain that the second power, which corresponds to a contribution that decreases with increasing centre-of-mass energy $\sqrt s$, is associated with $\rho, \omega, f_2, a_2$ exchange. Evidently the data require also the presence of a contribution that rises with $\sqrt s$. We say that this contribution is associated with pomeron exchange. Since the force that is responsible for $p$ and $\bar p$ interactions is QCD, we should like to try to understand

---

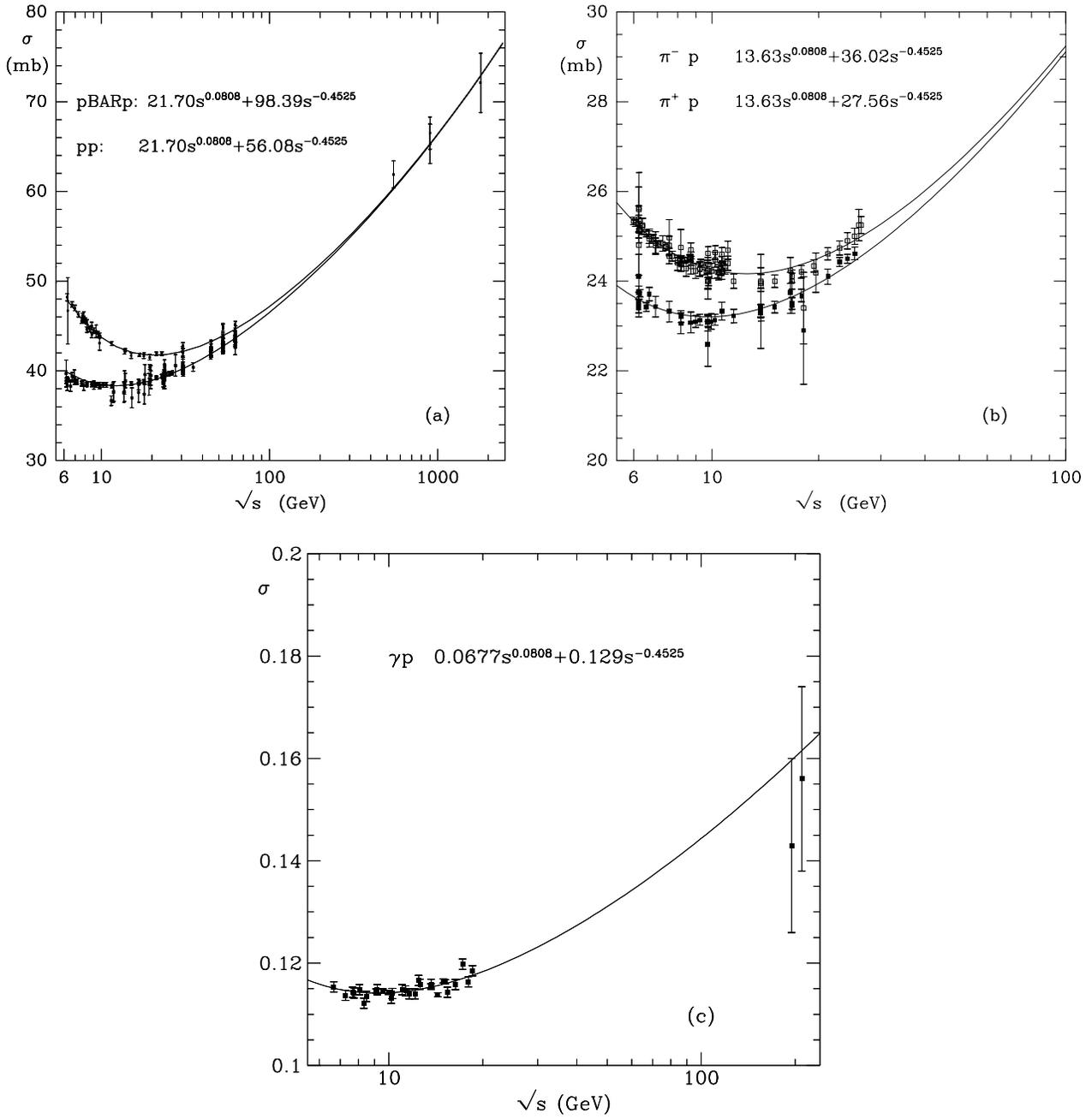

Figure 1: Total cross sections

how the pomeron is associated with QCD. The pomeron is supposed to have the quantum numbers of the vacuum, so it contributes equally to the $pp$ and $\bar{p}p$ total cross sections. This has been built into the fits, so that the two cross sections are described with 5 free parameters, two powers of $s$ and three multiplying coefficients. I should remark, for completeness, that the $\sqrt{s} = 1800$ GeV point shown in figure 1a is the measurement of the E710 experiment at the Tevatron[4]; the other experiment, CDF, reports a significantly larger value[5].

Having determined the two powers of $s$ from the $pp$ and $\bar{p}p$ data, we use the same powers for subsequent fits[3]. Figure 1b shows the $\pi^{\pm}p$ total cross sections. Again the pomeron-exchange term is contrained to have the same coefficient for each of these two cross sections, so this is a fit with 3 free parameters for the two cross sections. It is significant that the ratio of the strengths of pomeron exchange in $\pi p$ and $pp$ scattering is

$$\frac{13.6}{21.7} \approx \frac{2}{3}$$

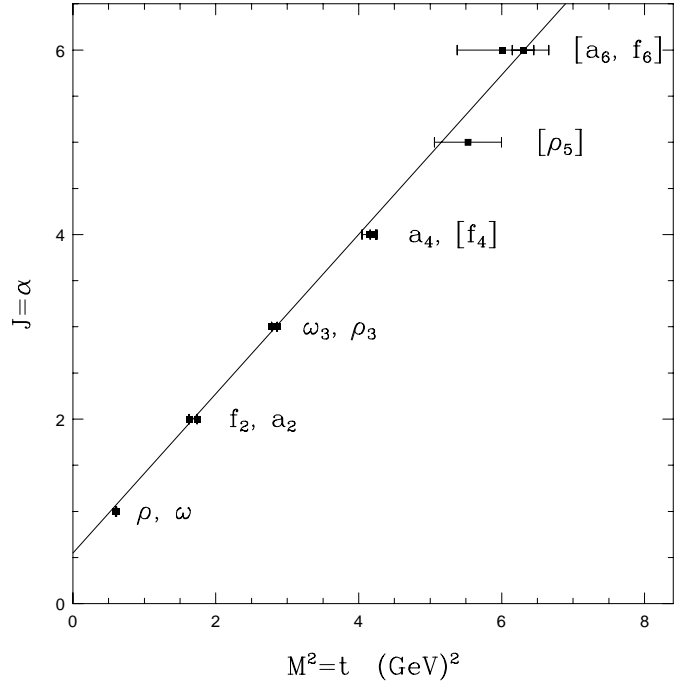

Figure 2: Regge trajectory – particle spins $\alpha$ plotted against their squared mass $t$

This is an indication that the pomeron couples to the valence quarks in the hadrons concerned (the proton has three, while the pion has only two). This *additive quark rule* is supported by other data[6] and is surprising: one might have expected that in high-energy scattering the notion of valence quarks was relevant only for hard interactions, not for soft pomeron exchange.

Figure 1c shows the real-photon–proton total cross section. Again the fit uses the same two powers; their multiplying coefficients are determined rather well because the low-energy data have such small error bars, and so allowed a prediction[7] for HERA energies, which are 10 times greater, with an error of $\pm$ a few $\mu$b.

## 3. REGGE THEORY

Figure 2 shows a plot of the spins $\alpha$ of the particles $\rho, \omega, f_2, a_2$ and their excitations, against their squared masses $t$. The particles in square brackets are listed in the data tables, though there is some uncertainty about them. The straight line has equation

$$\alpha(t) = 0.55 + 0.86t \qquad (1)$$

The line is extrapolated down to negative $t$, so that $t$ may then be regarded as a momentum-transfer variable. Then the exchange of all the particles associated with $\alpha(t)$ (see figure 3) gives to any elastic scattering amplitude a behaviour at high centre-of-mass energy $\sqrt{s}$

$$T(s,t) \sim \beta(t) s^{\alpha(t)} \xi_{\alpha(t)} \qquad (2a)$$

Here $\beta(t)$ is an unknown real function, while

$$\xi_{\alpha(t)} = \begin{cases} e^{-\frac{1}{2}i\pi\alpha(t)} & C = +1 \\ ie^{-\frac{1}{2}i\pi\alpha(t)} & C = -1 \end{cases} \qquad (2b)$$

where $C$ is the $C$-parity of the exchanged particles. Thus the "Regge trajectory" $\alpha(t)$ determines both the power of $s$ and the phase.

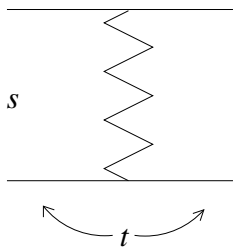

Figure 3: Regge exchange

There have been measurements of the phases of the $pp$ and $\bar{p}p$ elastic scattering amplitudes at zero momentum transfer $t$. Data up to the maximum CERN ISR energy, $\sqrt{s} = 63$ GeV, agreed well with the expectation from (1). A measurement by the UA4 collaboration at the CERN $\bar{p}p$ collider seemed to find that the agreement did not survive at higher energy, but then the E710 experiment at the Tevatron found that, at still higher energy, all is well. UA4 has repeated its measurement, and now concludes that the theoretical expectations are, after all, verified also at the CERN collider[4,8].

The optical theorem relates any total cross section to $s^{-1}$ times the imaginary part of the corresponding forward elastic scattering amplitude. The difference between $pp$ and $\bar{p}p$ scattering receives contributions only from $C = -1$ exchange in the $t$ channel, that is from $\rho$ and $\omega$ exchange. So we expect the difference between these total cross sections to behave as $s^{0.55-1} = s^{0.45}$. This feature is found in the fits of figure 1a. The average of the two cross sections receives only contributions from $C = +1$ exchange. We expect to see the same power of $s$ coming from the $f$ and the $a$, because according to figure 2 they lie on the same trajectory $\alpha(t)$ as the $\rho$ and the $\omega$, but evidently we need something else in order to reproduce the rise in the data at large $s$. The curves in figure 1 have this rising component, the pomeron term, as a simple power.

The obvious question is whether this simple pomeron-exchange power in fact results again from the exchange of a set of particles, in a similar way to the power associated with $\rho, \omega, f, a$ exchange. Theoretical prejudice suggests that if this is the case, the corresponding particles will be glueballs. As I shall explain, data for elastic scattering are well fitted[9] by supposing that the trajectory $\alpha(t)$ for pomeron exchange is straight, just like that for $\rho, \omega, f, a$:

$$\alpha(t) = 1 + \epsilon_0 + \alpha' t$$

$$\epsilon_0 = 0.086 \qquad (3a)$$

Whereas the slope of the $\rho, \omega, f, a$ trajectory shown in figure 2 is near to 1 GeV$^{-2}$, for the pomeron it is rather smaller:

$$\alpha' = 0.25 \text{ GeV}^{-2} \qquad (3b)$$

If this straight-line behaviour may validly be extrapolated to positive $M^2 = t$, then $\alpha(M^2)$ is equal to 2 at $M \approx 1900$ MeV. The line (3) is plotted in figure 4, which shows also a particle found by the WA91 collaboration at CERN[10]. WA91 have established that this particle is $2^{++}$, so it has the right quantum numbers to be the first particle on the pomeron trajectory. They say also that it is a glueball candidate, because they find it is produced isolated in rapidity, away from the other final-state particles (it decays to $\pi^+ \pi^- \pi^+ \pi^-$). This is the first hint, then, that we may make for pomeron exchange a plot just like figure 2, but with glueballs instead of quarkonium states.

The reason that $\epsilon_0$ in (3) takes a slightly higher value than the 0.0808 in the fits of figure 1 is that in these fits it is an effective power, which in fact is supposed to decrease (albeit very slowly) as $s$ increases. This variation is supposed to result from a a mixture of terms. At present energies the main one is single-pomeron exchange, which according to (3) behaves like $s^{0.086}$. The power closer to 0.08 is obtained by adding in double-pomeron exchange, which is negative and at CERN $\bar{p}p$ collider energy decreases the total cross-section by probably about 10%. At higher energies the contribution from double-pomeron exchange will become relatively larger. Eventually, one will also have to take account of the exchange of more than two pomerons (though up till now this seems to be unimportant), until at asymptotic energies an effective power $s^\epsilon$ is not a good representation and instead the Froissart bound[2] $\log^2 s$ is saturated.

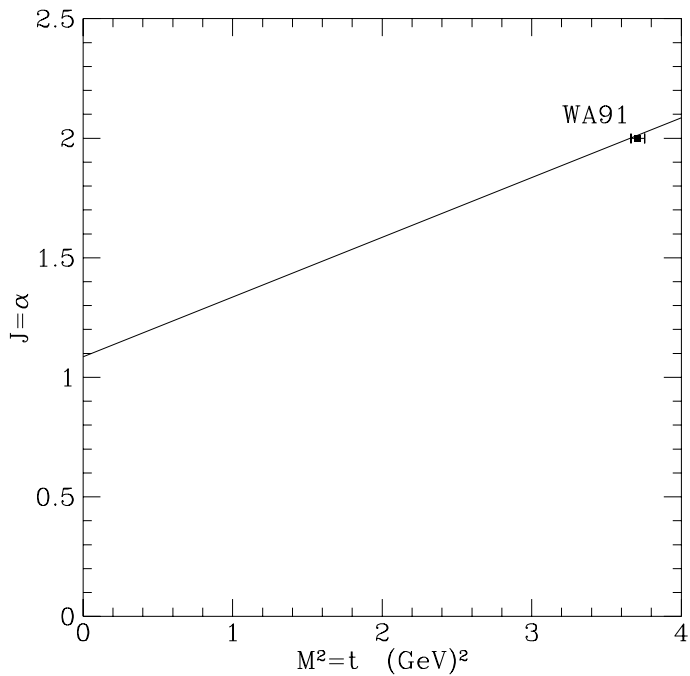

Figure 4: Pomeron trajectory with glueball candidate from the WA91 experiment[10]

I should say that not all authors agree that two-pomeron exchange is small[11]. Unfortunately, we are not able to calculate how strong two-pomeron exchange actually is, but must extract it from data. Those who believe it is large must also assume that the power $\epsilon_0$ corresponding to single exchange is somehat larger than given in (3). However, Donnachie and I have fitted[7, 9] a large quantity of data with the choice (3) and the assumption that two-pomeron exchange is small.

Notice that the same steadily-rising component $s^\epsilon$ is present in the data of figure 1a all the way from $\sqrt{s} = 5$ GeV to $\sqrt{s} = 1800$ GeV. This is remarkable, when one realises that all sorts of things are produced at 1800 GeV, particularly jets of not too high $p_T$ – known as minijets – that cannot be produced at 5 GeV. The total cross-section does not notice this new production and is smooth: as new final states come in, old ones are correspondingly reduced.

There is good experimental evidence, then, for the existence of the pomeron. This is the soft pomeron: it is nonperturbative in origin, has $\epsilon$ close to 0.08, and has been studied for more than 3 decades. Recently, the data from HERA for the small-$x$ behaviour of $\nu W_2$ have given a hint of the existence of another pomeron, the hard pomeron. This is perturbative and is supposed to have a much higher value of $\epsilon$, maybe even as large as 0.5. I shall discuss this later.

## 3. ELASTIC SCATTERING

When one considers all the available data for total cross-sections, elastic scattering and diffraction dissociation, the following phenomenological facts about the pomeron become apparent[7, 9]

- It couples to single quarks
- It yields simple power behaviour.
- It is rather like a $C = +1$ isoscalar photon

The first of these properties is tested not only by the validity of the additive quark rule for total cross sections, but also more directly in ISR data[12] for certain diffraction dissociation processes.

The property that the pomeron resembles a $C = +1$ isoscalar photon is tested in elastic scattering[9]. Just as the coupling of the photon to a proton involves a Dirac form factor, so will the coupling of the

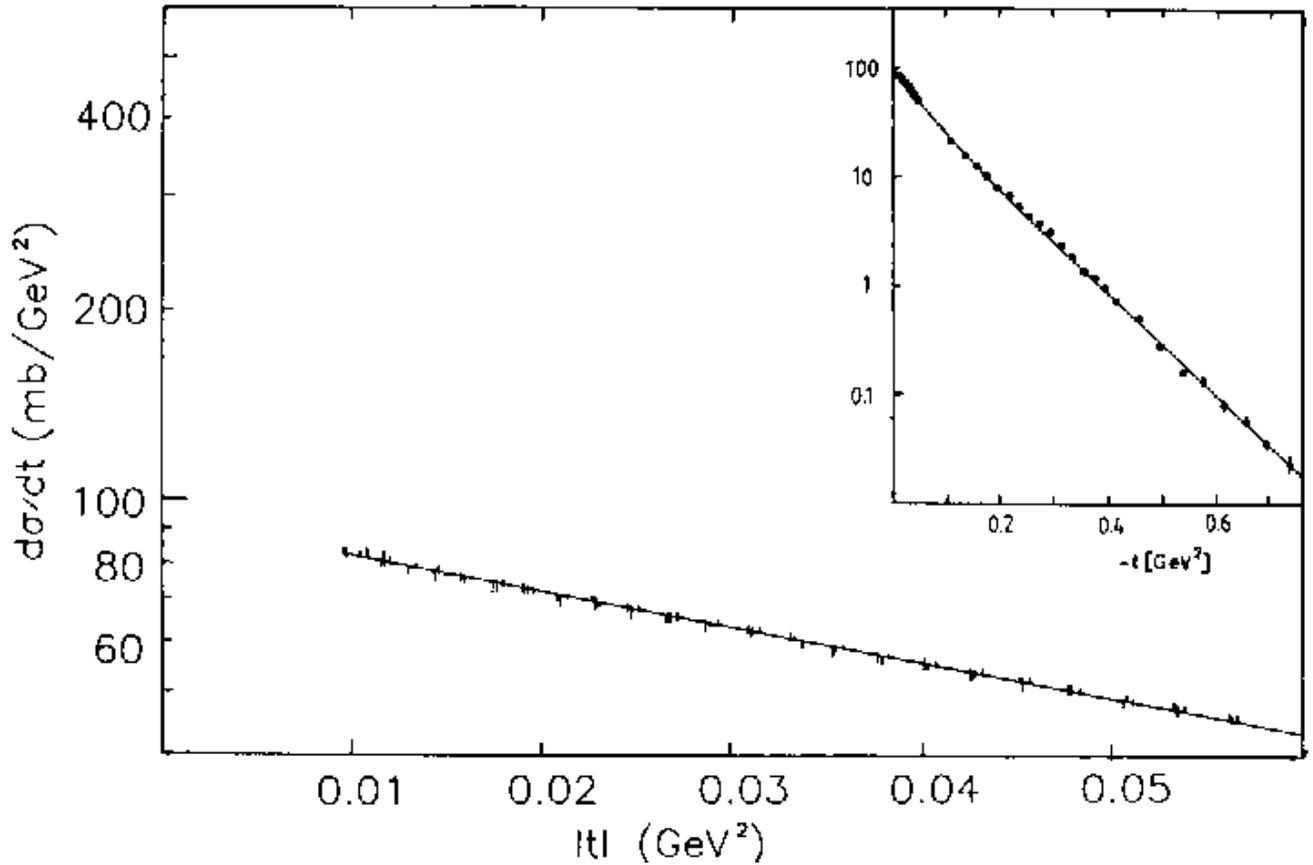

Figure 5: $pp$ elastic scattering at $\sqrt{s} = 53$ GeV

pomeron. So single pomeron exchange contributes to $pp$ or $\bar{p}p$ elastic scattering

$$\frac{d\sigma}{dt} = \text{constant } [F_1(t)]^4 s^{2\alpha(t)-2} \tag{4}$$

For want of any better knowledge we take the same isoscalar form factor $F_1(t)$ as has been measured in $eN$ scattering, and this seems to work well:

$$F_1(t) = \frac{4m^2 - 2.8t}{4m^2 - t}\left(\frac{1}{1 - t/0.7}\right)^2 \tag{5}$$

This corresponds to dipole $G_M$, and $G_M/G_E$ scaling. (There is no contribution from the other Dirac form factor $F_2(t)$, because this form factor is small in the isoscalar channel. This corresponds to the property that pomeron exchange does not flip the nucleon helicity[2].) In order to compare with data we have to add in to (4) the correction corresponding to two-pomeron exchange, which I have said is relatively small at small $t$, though less so at larger $t$, particularly at higher energies. Comparing the result with very-small-$t$ data at $\sqrt{s} = 53$ GeV determines $\alpha'$, and we then have an excellent fit to all ISR data out to values of $|t|$ of about 0.7 GeV$^2$: see figure 5. The same fit[9], with no adjustment since 1985 to the small number of parameters, correctly describes the recent Tevatron data[13] (figure 6). This is something of a triumph for Regge theory, which correctly predicted the change of exponential forward slope of about 3.5 units compared with ISR energies. Single-pomeron exchange gives a change of slope equal to $\Delta(2\alpha' \log s)$, though there is a small correction because of the two-pomeron exchange.

One may consider also other elastic-scattering processes: for example, proton-deuteron scattering data are described well by a similar analysis[7], involving now the elastic form factor of the deuteron extracted from elastic electron-deuteron scattering. Likewise, the quasielastic process $\gamma p \to \rho p$, which through vector dominance is related to elastic $pp$ scattering, also shows[14] a steepening of the forward peak with increasing energy.

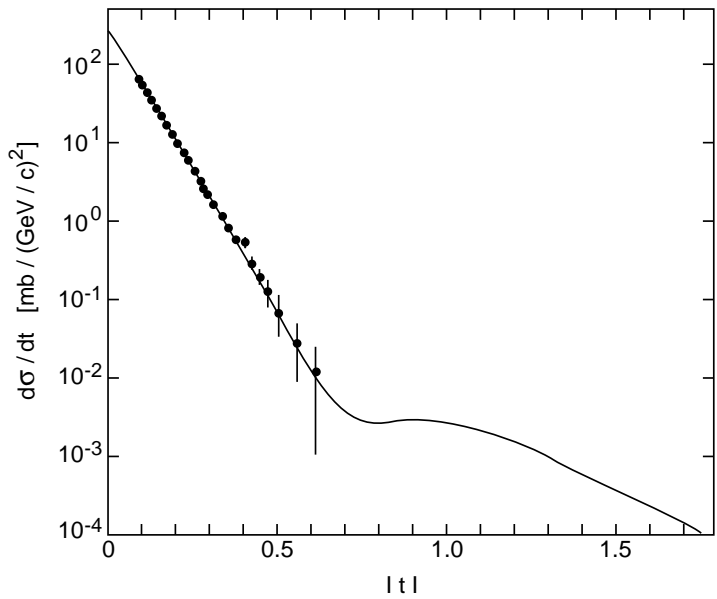

Figure 6: $\bar{p}p$ elastic scattering at $\sqrt{s} = 1800$ GeV

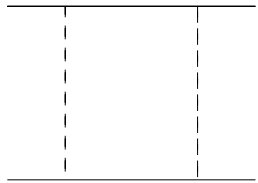

Figure 7: Exchange of a gluon pair between quarks

## 4. SEMIHARD PROCESSES

I have said that pomeron exchange may be glueball exchange. Whether or not this is true, it has long been believed that the simplest approximation to pomeron exchange is the exchange of a pair of gluons. (At least two gluons are needed, in order to yield colour-singlet exchange.) I have explained that the exchange takes place between quarks (figure 7).

Gluons are confined. This means that their propagator $D$ should not have the pole at $k^2 = 0$ that is there in perturbation theory: it is removed by nonperturbative effects. So it is possible that the integral

$$I = \int_{-\infty}^{0} dk^2 [\alpha_S D(k^2)]^2 \qquad (6a)$$

can be convergent. If it is, the two-gluon model for pomeron exchange gives[15] a total cross section equal to a constant times $I$. It also makes the two gluons together couple to the quarks effectively like a $C = +$ photon, as experiment seems to favour. What the simple model does not explain is the power $s^\epsilon$; this has to be included by hand.

Define now[15] the length $a$ by

$$\frac{1}{a^2} = I^{-1} \int_{-\infty}^{0} dk^2 (-2k^2) [\alpha_S D(k^2)]^2 \qquad (6b)$$

Because $1/a$ is a nonperturbative quantity, one would expect its value to be about 1 GeV; lattice calculations[16] seem to support this. One may think of $a$ as the confinement length, the greatest distance that confinement allows a gluon to travel. Assuming that the 1 GeV value is correct, then $a^2 \ll R^2$, where $R$ is the radius of a nucleon or a pion. Then one may show that the simple two-gluon-exchange model reproduces[15] the additive-quark rule for total cross sections.

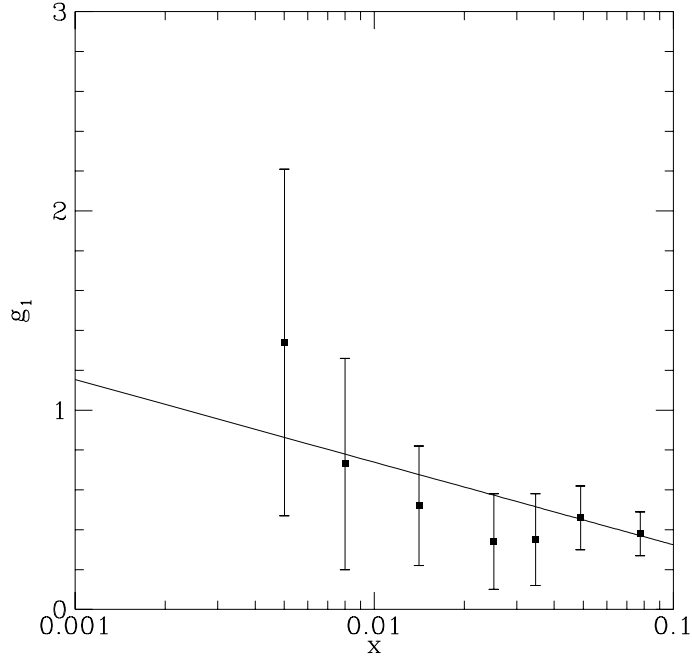

Figure 8: The structure function $g_1$ of the proton – data from the SMC experiment[22] with the curve (9)

The same simple model allowed a successful prediction[17] for the low-energy data obtained by the NMC collaboration[18] for the process

$$\gamma^* p \to \rho p$$

measured in inelastic muon scattering. The experiment is difficult: NMC found it was necessary to take the greatest care to ensure that the data were not contaminated by events in which the proton breaks up[19]. In the simple model, it is again necessary to include in the amplitude by hand a factor $(W^2)^{\alpha(t)}$, which makes the total cross section for the process rise by a factor close to 2 between $W = 10$ and 100 GeV. This factor appears naturally in a simple generalisation of the model[20], which relates the $t = 0$ amplitude to the gluon structure function of the proton. This also raises the interesting possibility that the increase with energy may be more rapid than simple soft pomeron exchange predicts, maybe even revealing the hard pomeron.

The model can also be applied to the small-$x$ behaviour of the polarised structure function $g_1$ of the proton. When two gluons couple to a quark, there is a $\gamma$ matrix for each coupling, and one for the quark propagator between them (see figure 7). There is a simple identity:

$$\gamma^\mu \gamma^\alpha \gamma^\nu = S^{\mu\alpha\nu} + A^{\mu\alpha\nu}$$
$$S^{\mu\alpha\nu} = g^{\mu\alpha}\gamma^\nu + g^{\nu\alpha}\gamma^\mu - g^{\mu\nu}\gamma^\alpha$$
$$A^{\mu\alpha\nu} = i\epsilon^{\mu\alpha\nu\rho}\gamma^5\gamma_\rho \qquad (7)$$

It is the $S$-term that is used to project out the pomeron part of the two-gluon exchange; it is insensitive to the spin state of the quark. The $A$-term yields an exchange whose sign depends on the quark spin state: it contributes to the small-$x$ behaviour of the polarised structure function $g_1$ of the proton. The behaviour is[21]

$$g_1 \sim C \left( 2 \log \frac{1}{x} - 1 \right) \qquad (9)$$

The constant $C$ has a value that is very model-dependent; a very crude model gives $C = 0.09$. The resulting fit is compared with data in figure 8. Note that a contributing Regge exchange, such as the $A_1$ with $\alpha(0) \approx -0.4$, would give a constant times $x^{-\alpha(0)}$. So the two gluons in this exchange efffectively have $\alpha(0) = 0$, rather than 1 as in pomeron exchange. It is not known whether $t$-channel iterations will keep the effective $\alpha(0)$ at 0, or will change its value, just as for pomeron exchange they are supposed to change 1 to 1.086.

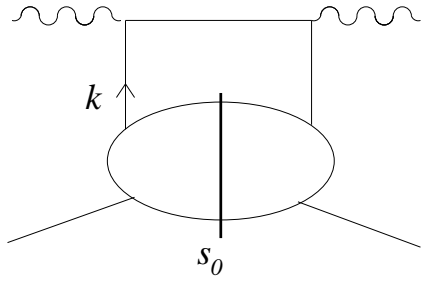

Figure 9: The covariant parton model[23]

I have suggested here that Regge theory is relevant for determining the small-$x$ behaviour of structure functions, and now explain this. Consider $\nu W_2$ at moderate $Q^2$-values, say 5 GeV$^2$, so that it may be reasonable to work to zeroth order in the perturbative QCD evolution. That is, use the parton model. There are two versions of the parton model, the naive parton model and a less naive one. The naive model requires one to choose a particular frame, the infinite momentum frame, and pretends that the intial parton is on-shell, $k^2 = 0$. The less naive model is covariant[23], and so does not require one to make a choice of frame; it is illustrated in figure 9. Here the parton that is extracted from the proton $p$ and is absorbed by the $\gamma^*$ can have any 4-momentum $k$. One must integrate over $k$, but the range of integration is effectively confined to values that are not too large, because of the assumed properties of the lower bubble, which is the nonperturbative amplitude for extracting the parton from the proton. Simple kinematics show that the squared invariant mass $s_0 = (p-k)^2$ of the system of residual fragments of the proton is, at small $x$,

$$s_0 \sim -\frac{k^2 + k_T^2}{x} \qquad (10)$$

and so is large. But the lower bubble in figure 9 is a strong-interaction amplitude, and $s_0$ is the corresponding squared energy for that amplitude. According to Regge theory, it should behave like $s_0^{\alpha(0)}$. When this behaviour is inserted in the calculation of figure 9, one finds[23] that the contribution to $\nu W_2$ behaves as the simple power $x^{1-\alpha(0)}$.

We use the same powers as in the fits of figure 1. We also remember that, when $Q^2 \to 0$, $\nu W_2$ is related to the total cross section for real-photon/proton scattering:

$$\sigma^{\gamma p} = \frac{4\pi^2 \alpha_{EM}}{Q^2} \nu W_2 \bigg|_{Q^2 = 0} \qquad (11)$$

so that $\nu W_2$ vanishes linearly with $Q^2$. Figure 10 shows the fit[24]

$$\nu W_2 = 0.32 \, x^{-0.08} \left(\frac{Q^2}{Q^2 + a}\right)^{1.08} + 0.10 \, x^{0.45} \left(\frac{Q^2}{Q^2 + b}\right)^{0.55}$$

$$a = (750 \text{ MeV})^2 \qquad b = (110 \text{ MeV})^2 \qquad (12)$$

together with the small-$x$ NMC data[25] for $Q^2 < 10$. The fit (12) is a two-parameter fit: for each choice of the coefficients multiplying the two terms, $a$ and $b$ are fixed by the requirement that, through (11), we retrieve the fit in figure 1c to the real-photon data. No perturbative evolution has been included in the curves of figure 10.

If we fix $Q^2 = 8.5$ and extrapolate (12) down to smaller $x$, we obtain the curve that is shown with measurements from HERA in figure 11. Some theorists believe that the difference between the curve and the data can be accounted for simply by including ordinary perturbative evolution[26]. Others prefer a more interesting explanation: there is a new contribution, the exchange of the hard pomeron.

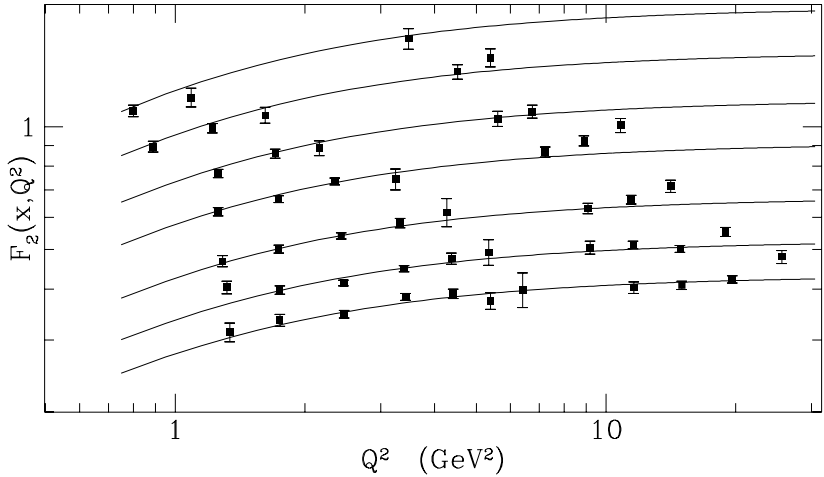

Figure 10: NMC data[25] with the curves (12). Reading from top to bottom, the $x$-values are 0.008, 0.0125, 0.0175, 0.025, 0.035, 0.05, 0.07. Curves and data have been scaled by a different factor at each value of $x$

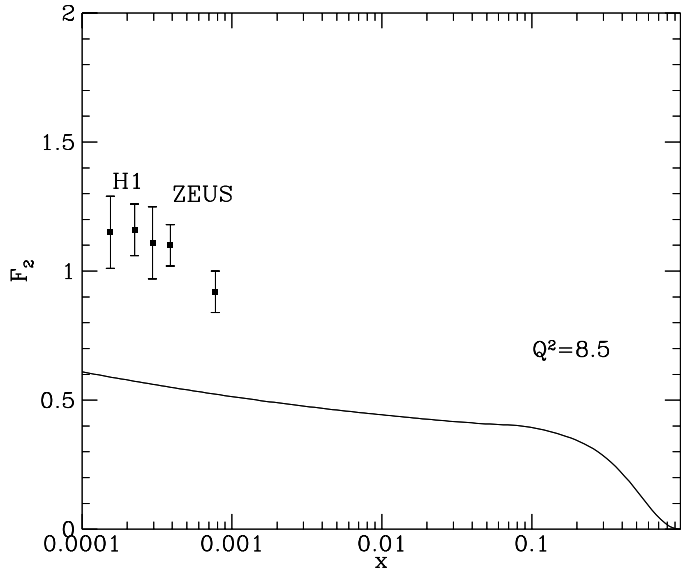

Figure 11: The curve (12) for $Q^2 = 8.5$ extrapolated to very small $x$, with preliminary HERA data

## 5. THE HARD POMERON

The hard pomeron is purely perturbative and is obtained by solving the BFKL equation[1]. This equation is derived by considering some Feynman graph $M$ (figure 12a) that contributes to the small-$x$ behaviour of $\nu W_2$ and then adding in an extra gluon, as for example in figures 12b and 12c. The sum of such insertions is found to have the structure $K \otimes M$, where $K$ is a certain function of $x$ and transverse momentum $k_T$, known as the Lipatov kernel. The convolution operation $\otimes$ includes an integration over $x$ and $k_T$. Summing over an infinite number of gluon insertions gives the small-$x$ behaviour of $\nu W_2$ as the solution to the integral equation

$$T = T_0 + K \otimes T \qquad (13)$$

Here $T_0$ is whatever contribution one starts with, before adding in gluon insertions. If all values of $k_T$

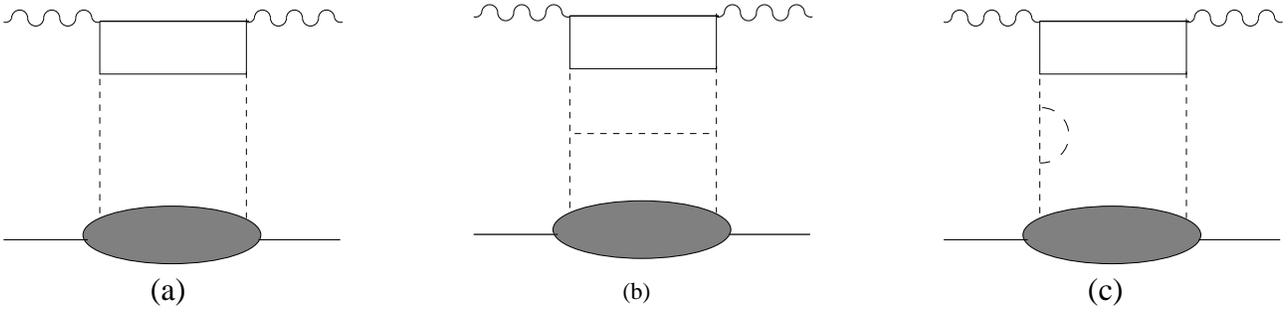

Figure 12: (a) A Feynman graph that contributes to $\nu W_2$ at small $x$, with examples of the addition of (b) a real and (c) a virtual gluon

are included in the integration, the solution behaves as $x^{-N}$, with

$$N = \frac{12\alpha_S}{\pi} \log 2 \qquad (14)$$

It is conventional to use here a value of $\alpha_S$ close to 0.2, so that $N \approx 0.5$. Compare this with 0.08 from soft pomeron exchange!

This raises two immediate questions:

(1) How does the hard perturbative pomeron relate to the soft nonperturbative one?

(2) It is surely incorrect to integrate over all values of $k_T$: for $k_T^2$ too small it is not valid to use the perturbative propagator, while values that are too large would violate energy conservation.

Collins and I constructed[27] a model that explored these two questions. Our model is less than perfect, but it suggests that the two pomerons should be added. This would imply that in figure 11 the hard-pomeron contribution is the difference between the data and the curve. As for the second question, we found that simply imposing a lower limit on the allowed values of $k_T^2$ in the integration does not change the value of the effective power $N$ given in (14), but imposing also an upper cut-off does have an effect. The value of $N$ then depends on the ratio $R$ of the upper and lower $k_T$ cut-offs. For $R = 100$, which is appropriate for the small-$x$ data in figure 11 under the assumption that the lower cut-off should be about 1 GeV, $N$ is reduced from 0.5 to 0.4. This has been checked by subsequent work[28], but in reality the effects of energy conservation will be very much more severe. Simply imposing an upper cut-off on $k_T$ guarantees that no final parton takes more than the total available energy, but rather one needs to ensure that the *sum* of the final-state energies is constrained. This has not been done. The effective value of $N$ will also be reduced significantly by multiple hard-pomeron exchange corrections to the single exchange. If the single exchange behaves like $x^{-N}$, the double exchange is a negative correction that behaves[2] as $x^{-2N}$. Furthermore, the BFKL equation at present only takes account of leading $\log x$ terms. Nonleading terms need to be included, if only to join up with the ordinary perturbative evolution at larger $x$ values, and this again[29] is likely to reduce the effective value of $N$.

It is clear from what I have said that a lot remains to be understood about hard pomeron exchange.

## 6. RAPIDITY GAP PHYSICS

Rapidity gap physics has created a lot of interest at HERA. The corresponding physics at hadron colliders has been well studied and is known as diffraction dissociation. Diffraction dissociation is the name given to inelastic events in which one of the initial hadrons changes its momentum by only a very small amount. In doing so, it 'radiates' a pomeron. The other initial hadron, or photon, is hit by the pomeron and breaks up into a system $X$ of hadrons (figure 13). The UA8 experiment at the CERN $\bar{p}p$ collider[30] has measured the angular distribution of the energy flow of the particles that make up the system $X$, in its rest frame, that is in the centre-of-mass frame of the pomeron-hadron collision. This is shown in figure 14, for events where the pomeron takes only a fraction 0.006 of the momentum of the initial hadron from which it was radiated. (This corresponds to an invariant mass of 50 GeV for the system $X$.) Notice the vertical logarithmic scale: there is a huge forward peak. The pomeron has hit the other initial hadron hard and knocked most of its fragments forward. In this respect it behaves as if it were itself a hadron, or a photon.

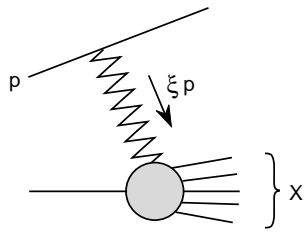

Figure 13: Diffraction dissociation

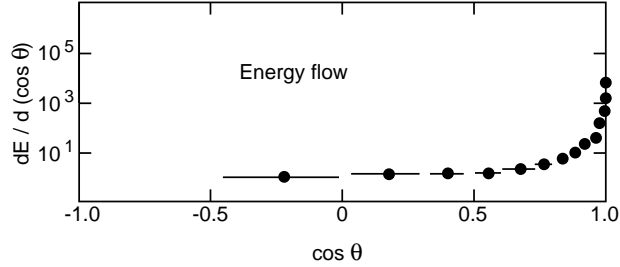

Figure 14: UA8 data for angular distribution of the energy flow in pomeron-proton collisions. The direction $\theta = 0$ is that of the incoming pomeron.

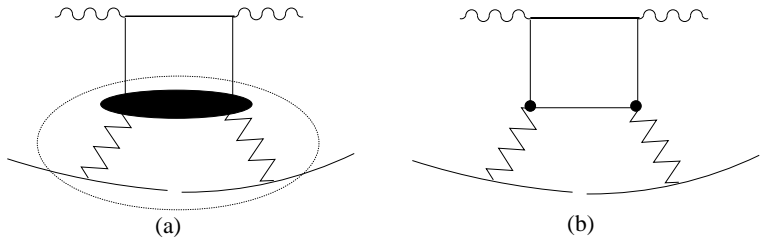

Figure 15: Diffractive deep inelastic scattering: (a) a part of figure 9, with (b) a simple model for the central bubble

In fact, as I have already said, it is useful to think of the pomeron as resembling a $C = +1$ isoscalar photon. One cannot take this analogy too far; for example, there is no such thing as a pomeron *state*, the pomeron can only be exchanged. However, one can define its structure function[31].

This is measured from events where the system $X$ has resulted from a hard collision, for example when it contains high-$p_T$ jets. UA8 has verified that this does occur[30]. At HERA, the most direct way to measure the pomeron structure function is simply to measure at high $Q^2$ the part of $\nu W_2$ corresponding to events in which there is a very fast proton in the final state. In terms of the covariant parton model, this corresponds to selecting events where the system $s_0$ in figure 9 includes a very fast proton. This is a leading-twist effect: it corresponds to picking out a particular part of the lower bubble in the figure, so that figure 9 becomes figure 15a. When there is a very fast proton, kinematics requires that there be a rapidity gap between it and the other final-state hadrons, though of course the presence of a rapidity gap is not by itself a sufficient indication that there is a fast proton. Given that, as we have seen, the pomeron couples to quarks similarly to a photon, it is natural to suppose that an important part of figure 15a is the simpler diagram shown in figure 15b where, as for the photon, the pomeron structure function is modelled by a simple box diagram. From this diagram, we predicted[32] that, for each light quark and antiquark to which the pomeron couples, its structure function is

$$xq_{\text{pomeron}} = \tfrac{1}{3}C\pi x(1-x) \tag{15}$$

where $C \approx 0.2$. Just as the photon structure function has a piece at small $x$ that is calculated from vector dominance, so the pomeron structure function also has an additional piece, but this is important only for very small $x$ and for most purposes (15) is sufficient[32]. Donnachie and I obtained this form of the structure function in a phenomenological model for the pomeron. We introduced a form factor into the coupling of the pomeron to the quarks, which is one reason why the $x$-dependence (15) is not the same as for the photon structure function. Diehl[33] has derived a similar form from the nonperturbative-gluon-exchange model of the soft pomeron, where effectively such a form factor emerges from cancellation among different diagrams, while Nikolaev and Zakharov[34] have similarly calculated the structure function of the pomeron, though they used perturbative gluon propagators.

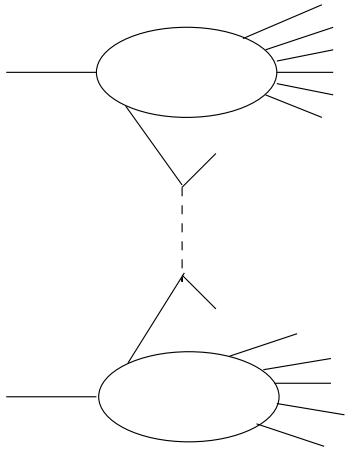

Figure 16: Hard scattering mechanism

The experimental situation is not yet clear. HERA probes the quark structure function of the pomeron directly in the large-rapidity gap events, but until the in-beam-pipe proton detector makes it possible to measure in what fraction of these the proton remains intact one can only say that (15) seems at least to be consistent with the data. However, CDF has measured[35] what fraction of $W$-production events at the Tevatron have a very fast proton in the final state, and find it to be extremely small, which may be difficult to reconcile with the HERA data since one would have thought that the same structure function is being measured. UA8 data on diffractive production of high-$p_T$ jets[30] are again consistent with (15), but it is not possible from the UA8 data to determine whether it is the quark structure function of the pomeron that is responsible for their events, or a gluon structure function. A previous experiment by UA1, which found evidence for the diffractive production of heavy flavours, suggests that there may be a significant gluon structure function[36].

Further experimental work will sort out the confusion. There also needs to be more theoretical work, particularly to understand the gluon structure function, and also the $Q^2$ evolution of both structure functions.

An obvious question is how to decide which of the two pomerons is being studied in a given experiment, if one is not able to measure its Regge trajectory $\alpha(t)$. The theoretical position is confused, because although the hard pomeron is supposed to be purely perturbative in origin, the soft pomeron is a mixture of perturbative and nonperturbative effects.

To explain this, let me first go back to the total cross section for real-photon–proton collisions. There were predictions that far exceeded the value measured at HERA. These were obtained by a standard hard-scattering calculation (figure 16) of $d\sigma^{\text{PAIR}}/dp_T$, the inclusive cross section for the production of a pair of jets of equal and opposite transverse momentum $p_T$. This was integrated down to some $p_T^{\min}$, chosen to be in the range 1 to 2 GeV. (When $p_T$ is as small as this, though still large enough for a perturbative calculation to be presumed valid, the jets are known as minijets.) This is a correct calculation, provided one recognises that the result is not a contribution to the total cross section, but rather[37]

$$\int_{p_T^{\min}} dp_T \; \frac{d\sigma^{\text{PAIR}}}{dp_T} = \bar{n}\rho\sigma^{\text{TOT}} \tag{16}$$

Here, $\rho$ is the fraction of events that have a jet pair, while $\bar{n}$ is the average number of jet pairs produced in such events. If the total cross section is less than the integral (16) then since obviously $\rho < 1$ it must be that $\bar{n} > 1$. This is the case for the upper energy ranges of the data shown in figures 1a and 1c,

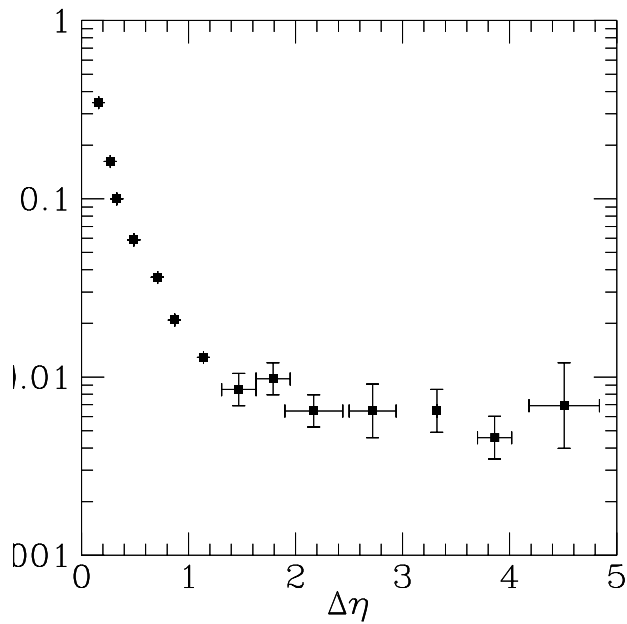

Figure 18: D0 data for fraction of jet-production events that have a rapidity gap between the jets, plotted against the rapidity separation of the jets

where there is copious production not only of minijets but also presumably of charm. Soft pomeron exchange, corresponding to a rising power $s^{0.08}$, describes the data well and therefore includes this minijet and charm production at the higher energies. It does not make sense to say that minijet and charm production *cause* the rise in the total cross section. As can be seen in the fits of figure 1, the rising component $s^{0.08}$ of the total cross section is present already at energies far too low for there to be any production of minijets or charm – according to the figures, the same rising term $s^{0.08}$ is present all the way from $\sqrt{s} = 6$ Gev to 1800 Gev; indeed[24], for $\gamma p$ scattering the fit is good even down to 2 GeV.

The obvious question, to which neither theory or experiment yet provides an answer, is whether this situation will persist to higher and higher energies, or whether eventually perturbative effects take over to such an extent that there is a transition from soft pomeron exchange to hard pomeron exchange.

An interesting suggestion[38] is that it may be possible to seek out the hard pomeron by looking for the production of a pair of jets with a large rapidity gap between them. In the central hard gluon exchange in figure 16 a colour string is created between the two emerging quark jets, and according to our understanding this leads at the subsequent hadronisation stage to the creation of particles intermediate in rapidity between the two jets. However, if instead the single hard gluon is replaced with a pair, the pair can form a colour singlet and so there need then be no colour string between the two jets. This does not guarantee that there will be a rapidity gap between the jets, because there is a colour string between the two systems of residual fragments of the hadrons, at the top and bottom of the diagram. It is not possible to estimate reliably how frequently the hadronisation asscociated with this will leave a rapidity gap[39], but there have recently been measurements of jet production with a rapidity gap by the D0 experiment at the Tevatron[40]. Figure 17 shows the fraction of jet production events that do have a gap between the two jets, plotted against the rapidity difference $\Delta\eta$ between the jets. The flattening off at large $\Delta\eta$ is certainly suggestive of two-gluon exchange. If this exchange is the beginning of hard pomeron exchange, one would expect the fraction even to increase again at larger $\Delta\eta$.

## 7. CONCLUSIONS

The soft pomeron has been familiar now for nearly 35 years. It has very simple phenomenology and has allowed many successful predictions. Soft pomeron exchange seems to be nonperturbative gluon exchange, and there is a hint that it may actually be glueball exchange.

The hard pomeron, according to our theoretical understanding, is perturbative gluon exchange. There are hints that it may have been observed in experiments at HERA and at the Tevatron, though this is not yet sure.

If there are two pomerons, a key question is how do they coexist? Do they simply add, or does the soft one go smoothly over to the hard one as the effective $Q^2$ increases?

This research is supported in part by the EU Programme "Human Capital and Mobility", Network "Physics at High Energy Colliders", contract CHRX-CT93-0357 (DG 12 COMA), and by PPARC.

**REFERENCES**

1. E A Kuraev, L N Lipatov and V S Fadin, Sov Physics JETP 44 (1976) 443
   L V Gribov, E M Levin and M G Ryskin, Physics Reports 100 (1983) 1
2. P D B Collins, *Introduction to Regge Theory and High Energy Physics*, Cambridge University Press (1977)
3. A Donnachie and P V Landshoff, Physics Letters B296 (1992) 227
4. E710 collaboration: N Amos et al, Physical Review Letters 68 (1992) 2433
5. CDF collaboration: F Abe et al, preprint FERMILAB-PUB-93-234-E
6. H J Lipkin, Nuclear Physics B78 (1974) 381
7. A Donnachie and P V Landshoff, Nuclear Physics B244 (1984) 322
8. UA4 collaboration: C Augier et al, Physics Letters B316 (1993) 448
9. A Donnachie and P V Landshoff, Nuclear Physics B267 (1986) 690
10. WA91 collaboration: S Abatzis et al, Physics Letters B324 (1994) 509
11. E Gotsman, E M Levin and U Maor, Physical Review D49 (1994) 4321
    A Capella et al, preprint hep-ph/9405338
12. R608 collaboration: Physics Letters B163 (1985) 267 and B283 (1992) 155
13. E710 collaboration: N Amos et al, Phys Lett B243 (1990) 158
14. D Aston et al, Nuclear Physics B209 (1982) 56
    T J Chapin et al, Physical Review D31 (1985) 17
15. P V Landshoff and O Nachtmann, Z Physik C35 (1987) 211
16. A Di Giacomo and H Panagopoulos, Physics Letters B285 (1992) 133
17. A Donnachie and P V Landshoff, Nuclear Physics B311 (1989) 509
18. NMC collaboration: P Amaudruz et al, Z Physik C54 (1992) 239
19. J R Cudell, Nuclear Physics B336 (1990) 1
20. S J Brodsky et al, preprint hep-ph/9402283
21. S D Bass and P V Landshoff, preprint hep-ph/9406350
22. SMC collaboration: D Adams et al, Physics Letters B329 (1994) 399
23. P V Landshoff, J C Polkinghorne and R D Short, Nuclear Physics B28 (1970) 210
24. A Donnachie and P V Landshoff, Z Physik C61 (1994) 139
25. NMC collaboration: P Amaudraz et al, Physics Letters B295 (1992) 159
26. R D Ball and S Forte, preprint hep-ph/9409373
27. J C Collins and P V Landshoff, Physics Letters B276 (1992) 196

28  J R Forshaw, P N Harriman and P J Sutton, Nuclear Physics B416 (1994) 739

29  R K Ellis, Z Kunszt and E M Levin, Nuclear Physics B420 (1994) 517
    H Navelet, R Peschanski and S Wallon, preprint SACLAY-SPHT-94-012
    S Catani and F Hautmann, preprint hep-ph/9405388

30  UA8 collaboration: P Schlein, Nuclear Physics B (Proc Suppl) 33A,B (1993) 41

31  G Ingelman and P Schlein, Physics Letters B152 (1985) 256

32  A Donnachie and P V Landshoff, Phys Lett B191 (1987) 309 and Nuclear Physics B303 (1988) 634

33  M Diehl, preprint hep-ph/9407399

34  N N Nikolaev and B G Zakharov, Z Physik C53 (1992) 331

35  M G Albrow, preprint FERMILAB Conf-94/280-E

36  K Eggert, Proc 2nd International Conference on Elastic and Diffractive Scattering (Editions Frontières, 1988, ed K Goulianos)

37  M Jacob and P V Landshoff, Modern Physics Letters A1 (1986) 657
    A Donnachie and P V Landshoff, Physics Letters B332 (1994) 433

38  A H Mueller and H Navelet, Nuclear Physics B282 (1987) 727

39  J D Bjorken, Physical Review D47 (1993) 101

40  A Brandt, XVII International Conf on High Energy Physics, Glasgow